\documentclass[12pt]{article} 
\usepackage[dvips]{graphics}
\title{ Rotations and $e$, $\nu$ Propagators, Part III}  
\author{{\it Richard Shurtleff~}\thanks{affiliation and mailing 
address: Department of Mathematics and Applied Sciences, 
Wentworth Institute of Technology, 550 Huntington Avenue, 
Boston, MA, USA, ZIP 02115, telephone number: (617) 989-4338, fax 
number: (617) 989-4591 , e-mail address: shurtleffr@wit.edu}} 
\begin{document} 
          
\maketitle               
			\begin{abstract}  

In Parts I and II we showed that $e, \nu$ propagators can be derived from rotation invariant projection operators, thereby providing examples of how quantities with spacetime symmetry can be obtained by constraining rotationally symmetric objects. One constraint is the restriction of the basis; only two kinds of bases were considered, one for the electron and one for the neutrino. In this part, we find that, of a wide range of bases each consistent with the constraint process, only the two kinds of bases considered in Parts I and II give spacetime symmetric propagators. We interpret the result geometrically. The spinor representation is unfaithful in four dimensional Euclidean space which explains why spin 1/2 wave functions have four, not two, components. Then we show how a basis relates to two planes in four dimensional Euclidean space. A pair of planes spanning two or three dimensions does not allow spacetime symmetry. Spacetime symmetry requires two planes that span four dimensions.   

 
		\end{abstract}
\pagebreak

\section{Introduction} \label{intro} 

	We begin with rotations and vectors in four dimensional Euclidean space $E^{4}.$ With a two dimensional representation we should be able to represent vectors with two component quantities, the 2-vectors. But we can't. In four dimensional Euclidean space the two dimensional representation is unfaithful, one matrix represents two different rotations. We need two 2-vectors to describe the motion of one vector in $E^{4}.$ The geometry is discussed in Sec.~2 more fully but basically one 2-vector moves under rotations in one plane while the second 2-vector moves with rotations in a second plane. This may be why electron wave functions, which to begin with here are acted on only by rotations, have four components.  

	In Sec.~3 we consider bases whose upper 2-vectors are eigenvectors of rotations in one plane and the lower 2-vectors are eigenvectors in a second plane. In Sec.~2 the two planes are complementary, i.e. the two span all four dimensions, while in Sec.~3 we consider any two planes in the four dimensional Euclidean space. Eigenvectors of one plane are related to the eigenvectors of the second plane by a U(1)$\times$SU(2) transformation, in general. 

	In Sec.~4 the four projection operators for the general eigenvector basis are molded to yield two spacetime projection operators by the process developed in Parts I and II, \cite{partI} and \cite{partII}. The requirements of spacetime symmetry determine the U(1)$\times$SU(2) transformation relating the two rotation planes of the upper and lower 2-vectors. We find that only when the two planes for the basis span all four dimensions does the basis yield spacetime symmetric operators. These bases are just the ones considered in Parts I and II that turned out electron and neutrino propagators. Thus the process of constraining rotation-based quantities to make spacetime symmetric objects with a two dimensional representation can give only electron or neutrino propagators. Furthermore, from our viewpoint, having spacetime symmetry is equivalent to choosing complementary planes for rotating the upper and lower 2-vectors in the wave function.


\section{The Implied Geometry of Internal Spin Space} \label{geometry} 

	Why pairs of 2-vectors? Each electron or neutrino wave function has four components arranged in a pair of 2-vectors ( a chiral representation). Only the free neutrino is discussed in Part II, so agreement with the standard electroweak model would require that only two of the four components participate in interactions. 

	The explanation depends on the fact that the $2\times 2$ matrix representations of rotations are faithful in three dimensional Euclidean space $E^{3}$ but unfaithful in $E^{4}.$ For each of the two inequivalent representations of rotations in $E^{3},$ there is a one-to-one correspondence between matrices and rotations but in $E^{4}$ the correspondence is one-to-two. As we now show, to properly represent a vector quantity $\chi$ in $E^{4},$ one needs two 2-vectors so that the effects on $\chi$ of the two rotations can be distinguished. See Fig.~1 and 2. 

\begin{figure}[h] \label{fig1}	
\vspace{0in}
\hspace{0in}\includegraphics[0,0][288,288]{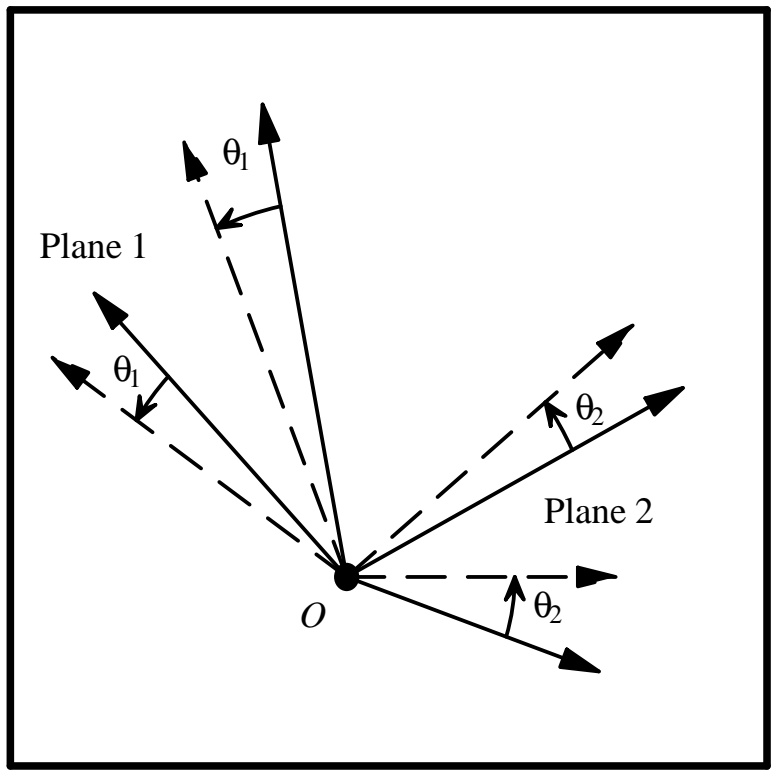}
\caption{ Vectors in planes 1 and 2 rotate. In four dimensional Euclidean space $E^{4},$ the rotations $r_{1}$ and $r_{2}$ are independent whenever planes 1 and 2 intersect only at the origin $O.$ In such cases, the $2 \times 2$ matrices $R_{1}$ and $R_{2}$ that represent the rotations must commute. Hence the matrices $R_{1}$ are the same matrices as the $R_{2}$s. Since one matrix represents two rotations, the representation is unfaithful.}
\end{figure}

	$E^{3}.$ A rotation $r$ in $E^{3}$ can be represented by a $2\times 2$ matrix $R,$ I(1), in either of two inequivalent matrix representations, also known as inequivalent `spinor' representations. Each representation is faithful, there is exactly one matrix $R$ for each rotation $r.$ Let $x$ be a 2-vector representing some vector quantity $\chi.$ Then applying the rotation $r$ with arbitrary rotation angles $\theta$ gives the evolution of the quantity, i.e. $r \chi.$ Matrix multiplication, $Rx$, gives the 2-vector that corresponds to $r \chi.$ Hence the evolution of the vector $\chi$ is well represented by one 2-vector $x.$

\pagebreak

\begin{figure}[h] \label{fig2}	
\vspace{0in}
\hspace{0in}\includegraphics[0,0][288,288]{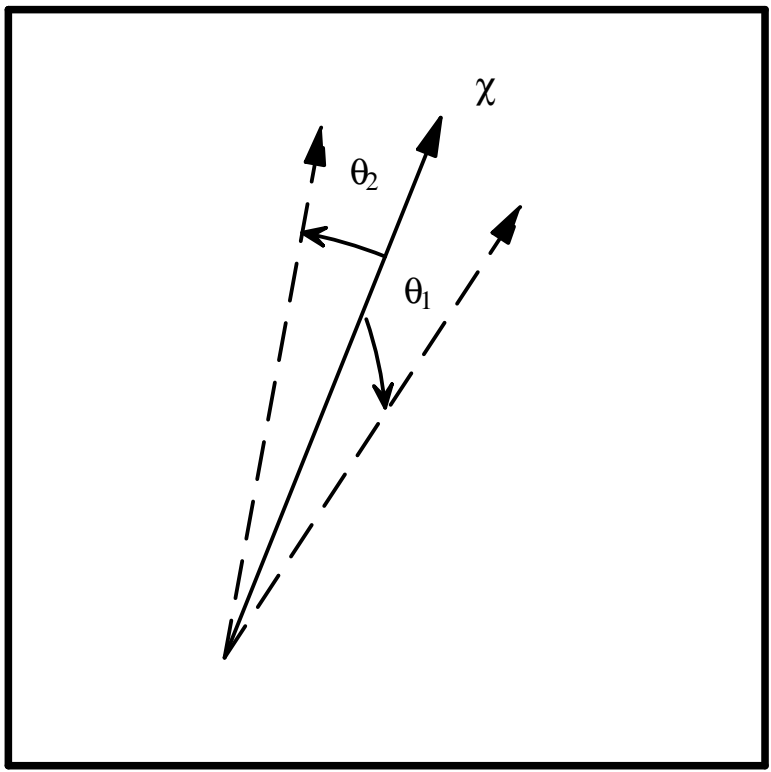}
\caption{ The vector $\chi$ reacts differently to rotations $r_{1}$ and $r_{2}.$ But when planes 1 and 2 intersect only at the origin, the same matrix $R$ represents both rotations. To keep track of what is happening to $\chi$ in two planes, we need two 2-vectors, thereby making a four component quantity $\psi.$ The upper 2-vector in $\psi$ represents $\chi$ rotating by $r_{1}$ and the lower represents rotation by $r_{2}.$ In Sec.~\ref{sectionspacetime} we show that spacetime can result only when the two planes span $E^{4}.$}
\end{figure}

	$E^{4}.$ In  four dimensional Euclidean space $E^{4}$ things are different, see Fig.~1. Given any 2-dimensional plane, plane 1, that contains the origin, there is a second plane, plane 2, that intersects 1 only at the origin. A rotation $r_{1}$ about the origin in plane 1 is independent of any rotation $r_{2}$ about the origin in plane 2, hence order is unimportant and $r_{1} r_{2} -$ $r_{2} r_{1}$ = 0. This implies that any matrix $R_{1}$ that represents $r_{1}$ must commute with the matrix $R_{2}$ that represents $r_{2}.$ For $R_{1}$ and $R_{2}$ this in turn means the unit 3-vectors $n^{k}$ in $R$ = $\exp{(i n^{k} \sigma^{k} \theta/2)}$ are equal, $n_{1}^{k}$ = $n_{2}^{k}.$ Thus $R_{1}$ and $R_{2}$ are the same matrix except with possibly different rotation angles $\theta_{1}$ and $\theta_{2}.$ This one-matrix-for-two-rotations association makes the $2 \times 2$ matrix representation `unfaithful' in $E^{4}.$

	When a 2-vector $x$ evolves by multiplying it by a rotation matrix $R$ with rotation angle $\theta,$ all we know is that a rotation $r_{1}$ through an angle $\theta_{1}$ and an independent rotation $r_{2}$ through an angle $\theta_{2}$ has taken place, where
\begin{equation}	\label{1+2}
\theta = \theta_{1} + \theta_{2}
\end{equation}
But the evolution of a vector quantity $\chi$ will in general depend on whether it is transformed via $r_{1}$ or $r_{2},$ see Fig.~2. Hence to keep track of the evolution of $\chi$ we need two 2-vectors, one for its evolution with rotations in plane 1, and one 2-vector for rotations in plane 2. 

	The spin 1/2 wave functions $\psi$ are just combinations of the two 2-vectors that represent vectors like $\chi$ in four dimensional Euclidean space $E^{4}.$  Rotation of the upper 2-vector in $\psi$ can take place in plane 1, while the rotations of the lower 2-vector take place in plane 2. 

	We can now answer the question posed at the start of the section. Each $\psi$ must have four components because the $2 \times 2$ matrix representations of rotations in $E^{4}$ are unfaithful and it takes two 2-vectors to represent one vector.

\section{An Eigenvector Basis with Independent Planes} \label{sectionortho}

	Up to this point in Parts I and II, the basis pairs have contained matched upper and lower 2-vectors, `matched' because they are eigenvectors $u^{+}$ or $u^{-}$ of the same rotation matrix $R,$ I(1) and I(3). We use eigenvectors because they give a phase factor when acted on by the rotation matrix and we use the phase factors in the definition of time and space. But, as just discussed, the rotation plane may not be the same for the upper and lower 2-vectors even when the same rotation matrix $R$ is applied to the upper and lower 2-vectors. What happens when the rotation matrices, and therefore the rotation planes, for the upper and lower 2-vectors are allowed to differ? 

	{\textit{Eigenvector Pairs}}. First we consider two general pairs. Let $\alpha,$ $\beta,$ $a,$ and $b$ be four 2-vectors arranged in two pairs,
\begin{equation}	\label{ab}
u^{\alpha}_{a} = \pmatrix{ e^{w_{1}/2} \alpha \cr e^{w_{2}/2} a } \hspace{0.5cm} u^{\beta}_{b} = \pmatrix{ e^{ w_{3}/2} \beta \cr e^{w_{4}/2} b } ,
\end{equation}
where we absorb phase factors in the 2-vectors thereby making the $w$s real and we absorb the normalization factors in the $w$s so that the 2-vectors are normalized to unity,
\begin{equation}	\label{normal0} 
\alpha^{\dagger} \alpha = 1 \hspace{1cm} \beta^{\dagger} \beta = 1 \hspace{1cm} a^{\dagger} a = 1 \hspace{1cm} b^{\dagger} b = 1 .
\end{equation}

	The two pairs (\ref{ab}) are orthogonal when they are `eigenvector pairs,' i.e. $\alpha$ and $\beta$ are eigenvectors of a rotation matrix $R$ and $a$ and $b$ are eigenvectors of a possibly different rotation matrix $R_{a}.$ To show this, note that since eigenvectors are orthogonal, I(7), we have
\begin{equation}	\label{normal2} 
\alpha^{\dagger} \beta = 0 \hspace{1cm} a^{\dagger} b = 0 
\end{equation}
and (\ref{ab}) then implies that $u^{\alpha}_{a},$ and $u^{\beta}_{b}$ are orthogonal,
\begin{equation}	\label{normal5} 
{u^{\alpha}_{a}}^{\dagger} u^{\beta}_{b} = 0.
\end{equation}
The inverse is not true. If the pairs (\ref{ab}) are orthogonal as in (\ref{normal5}) then it does not follow that $\{\alpha, \beta\}$ and $\{a, b\}$ are sets of eigenvectors and orthogonal.

	Hence we can associate planes with the upper and lower 2-vectors of eigenvector pairs. $\alpha$ and $\beta$ are eigenvectors for rotations in one plane and $a$ and $b$ are eigenvectors for rotations in a second plane. We may just as well take $\alpha$ and $\beta$ be the eigenvectors $u^{+}$ and $u^{-}$ for the rotation matrix $R$ in I(1) and I(3). We have
\begin{equation}	\label{alphabeta}
\alpha = u^{+} \hspace{0.5cm} \beta = u^{-} .
\end{equation}

	{\textit{U(1)$\times$SU(2) Theorem.}} One orthonormal set of eigenvectors is related to a second orthonormal set of eigenvectors by the action of a rotation matrix like $R$, I(1), and a phase factor, i.e. a U(1)$\times$SU(2) transformation. This is a well known assertion which is obtained in Appendix A for completeness. By (\ref{abu+u-}) we get
\begin{equation}	\label{abu+u-1}
 \pmatrix{ a  \cr b} = T \pmatrix{ u^{+}  \cr u^{-}} = e^{i\chi} [\tau^{4} \cos (\kappa/2) + i N^{k} \tau^{k} \sin (\kappa/2)]\pmatrix{ u^{+}  \cr u^{-}} = e^{i\chi} e^{i N^{k} \tau^{k} \kappa/2} \pmatrix{ u^{+}  \cr u^{-}},
\end{equation}
where $T$ is a phase factor times a rotation matrix, $\chi$ and $\kappa$ are real, $N^{k}$ is a unit 3-vector and the $2 \times 2$ matrices $\tau$ are the same as the $\sigma$ matrices in I(2). (We write a spin matrix as a $\tau$ when it is applied to a pair of 2-vectors and we write the spin matrix as a $\sigma$ when it is applied to the two components of one 2-vector.)

	The group U(1)$\times$SU(2) is the familiar gauge group for the electroweak interaction. It is important that a gauge group realize some special property of the system described. We have just found that U(1)$\times$SU(2) relates the eigenvectors of rotations in the upper 2-vector plane with those of the lower 2-vector plane. Put another way, the two pairs (\ref{ab}) are orthogonal for any values of $\chi,$ $N^{k},$ and $\kappa$ in $T,$ (\ref{abu+u-1}).

	{\textit{Basis.}} A basis contains four pairs, so we take two sets of eigenvector pairs like (\ref{ab}). We have
\begin{equation}	\label{uvabcd}
u^{+}_{a} = \pmatrix{ e^{d_{1}/2} u^{+} \cr e^{d_{2}/2} a } \hspace{0.5cm} u^{-}_{b} = \pmatrix{ e^{d_{3}/2} u^{-} \cr e^{d_{4}/2} b } \hspace{0.5cm} v^{+}_{a}= \pmatrix{ e^{i \alpha_{1}} e^{d_{5}/2} u^{+} \cr  e^{i \alpha_{2}} e^{d_{6}/2} a } \hspace{0.5cm} v^{-}_{b} = \pmatrix{ e^{i \alpha_{3}} e^{ d_{7}/2} u^{-} \cr e^{i \alpha_{4}} e^{d_{8}/2} b },
\end{equation}
where $\{u^{+}, u^{-}\}$ and $\{a, b\}$ are orthonormal sets of eigenvectors just as before and the phases $\alpha_{i}$ are arbitrary. 

	We assume the basis pairs are orthogonal. From ${u^{+}_{a}}^{\dagger} v^{+}_{a}$ = ${u^{-}_{b}}^{\dagger} v^{-}_{b}$ = 0, we get
\begin{equation}	\label{ortho78}
 \alpha_{1} - \alpha_{2} = \pi + 2n \pi \hspace{0.6cm} \alpha_{3} - \alpha_{4} = \pi + 2m \pi \hspace{0.6cm} d_{1} - d_{2} = -(d_{5} - d_{6}) \hspace{0.6cm} d_{3} - d_{4} = -(d_{7} - d_{8}) ,
\end{equation}
where $n$ and $m$ are integers. By orthogonality, (\ref{ortho78}), we can rewrite the basis (\ref{uvabcd}) as
$$u^{+}_{a} = e^{w_{1}/2}\pmatrix{ e^{w_{2}/2} u^{+} \cr e^{-w_{2}/2} a } \hspace{0.5cm} u^{-}_{b} = e^{w_{3}/2} \pmatrix{ e^{-w_{4}/2} u^{-} \cr e^{w_{4}/2} b } $$
\begin{equation}	\label{uvabcd2} 
v^{+}_{a}= e^{i\alpha} e^{w_{5}/2} \pmatrix{ e^{-w_{2}/2} u^{+} \cr  - e^{w_{2}/2} a } \hspace{0.5cm} v^{-}_{b} = e^{i \delta} e^{ w_{6}/2} \pmatrix{ - e^{ w_{4}/2} u^{-} \cr  e^{-w_{4}/2} b },
\end{equation}
where the $w$s are related to the $d$s, $w_{1}$ = $(d_{1} + d_{2})/2,$ $w_{2}$ = $(d_{1} - d_{2})/2,$ etc.

	As with previous bases we rotate both upper and lower 2-vectors so that the basis acquires a common phase. We get
\begin{equation}	\label{Rauv}
R_{a}^{+} u_{a}^{+} =  e^{+i\theta/2} u_{a}^{+} \hspace{0.5cm} R_{b}^{-} u_{b}^{-} =  e^{+i\theta/2} u_{b}^{-} \hspace{0.5cm} R_{a}^{+} v_{a}^{+} =  e^{+i\theta/2} v_{a}^{+} \hspace{0.5cm} R_{b}^{-} v_{b}^{-} =  e^{+i\theta/2} v_{b}^{-},
\end{equation}
where $R_{a}^{+}$ means applying $R$ to the upper 2-vector to generate the eigenvalue $\exp{(i\theta/2)}$ and $R_{a}$ to the lower 2-vector to generate the eigenvalue $\exp{(i\theta/2)}$, $R_{b}^{-}$ means applying $R^{-1}$ to the upper 2-vector to generate the eigenvalue $\exp{(i\theta/2)}$ and $R_{b}$ = ${R_{a}}^{-1}$ to the lower 2-vector to generate the eigenvalue $\exp{(i\theta/2)}.$  

	The rotated basis pairs (\ref{Rauv}) are not yet normalized to the same value. For example, we get	 
\begin{equation}	\label{uvortho}
 (R_{a}^{+} u_{a}^{+})^{\dagger} R_{a}^{+} u_{a}^{+} = {u_{a}^{+}}^{\dagger} u_{a}^{+} = 2 e^{w_{1}} \cosh w_{2},
\end{equation}
with similar expressions for the other normalization values. Normalization values are constrained by spacetime symmetry, as we find in the next section.

\section{Projection Operators and Space-time} \label{sectionspacetime}

	Any ordered set of four complex numbers can be written in terms of the basis pairs (\ref{Rauv}). Choose one such set and call it $\psi.$ We get
\begin{equation} \label{psi,ab}
 \psi = \alpha  R_{a}^{+} u_{a}^{+} + \beta R_{b}^{-} u_{b}^{-} + \gamma  R_{a}^{+} v_{a}^{+} + \delta R_{b}^{-} v_{b}^{-} = e^{i \theta/2}(\alpha  u_{a}^{+} + \beta u_{b}^{-} + \gamma  v_{a}^{+} + \delta v_{b}^{-}),
\end{equation}
where $\alpha$ = ${(R_{a}^{+} u_{a}^{+})}^{\dagger} \psi /(2 e^{w_{1}} \cosh w_{2}),$ etc.

	We now follow the procedure that gave projection operators I(22) and II(6) for the $uv$ basis in Sec.~4 of Part I and the $fg$ basis in Sec.~3 in Part II. We consider the basis-specifying parameters to be the $w$s in (\ref{uvabcd2}) and the $n^{k}$ of $u^{+}$ and $u^{-}.$ This implies that $a$ and $b$ are known in terms of $n^{k}$ via the U(1)$\times$SU(2) transformation (\ref{abu+u-1}). Hence when the as-yet-unknown delta function parameters $p^{\mu}$ are determined, they will be functions of $w$ and $n^{k},$ just as in Parts I and II.

	{\textit{Projection Operators}}. One can verify that the resulting projection operator for $R_{a}^{+} u_{a}^{+}$ is
\begin{equation}	\label{K(2,1)ab}
  K(2,1,R_{a}^{+} u_{a}^{+}) \gamma^{4} \equiv \int \frac{d^3 p^{\prime} }{(2 \pi)^3} \frac{1}{2 e^{{w_{1}}^{\prime}} \cosh w_{2}^{\prime}}e^{i \theta_{2}^{\prime}/2}  u_{a}^{+ \, \prime} {u_{a}^{+ \, \prime }}^{\dagger} e^{ -i \theta_{1}^{\prime}/2} ,
\end{equation} 
which gives
\begin{equation}	\label{K(2,1)psiab}
 \int d^{3} x_{1} K(2,1,R_{a}^{+} u_{a}^{+}) \gamma^{4} \psi(1) =  e^{i \theta_{2} / 2} (\alpha  u_{a}^{+}) = \psi(2)_{\beta = \gamma = \delta = 0},
\end{equation}
where $\psi(1)$ is $\psi$ with $\theta$ = $\theta_{1}$ and $ \theta_{i}^{\prime}/2 $ = $- \Delta_{i}^{\prime} +  {p^{k }}^{\prime} x_{i}^{k} .$ The projection operators for the other basis pairs are similar.

	{\textit{Time, Surface Integral}}. Following the steps leading to I(28) that give space-time symmetry in all parts except for the matrix part, we can write the action of the projection operator in near-space-time form. We get
\begin{equation}	\label{QEDnu3}
\psi(2)_{\beta = \gamma = \delta = 0} = i\int \frac{d^4 x_{1} d^4 p^{\prime} }{(2 \pi)^4} e^{-ip_{\mu}^{\prime}x_{2}^{\mu}} \; \frac{ (\pm m) {u_{a}^{+}}^{\prime} {{u_{a}^{+}}^{\prime}}^{\dagger}  \gamma^{4}}{{p^{4 \; \prime}}^2-{(m e^{{w_{1}}^{\prime}} \cosh w_{2}^{\prime})}^2  + i\epsilon} e^{ip_{\nu}^{\prime}x_{1}^{\nu}}  N_{\sigma} \gamma^{\sigma} \psi(1),
\end{equation}
where we introduce a fourth component of momentum and a fourth component of displacement, i.e. energy and time,
\begin{equation}	\label{p4t}
{p^{4}}^{\prime} \equiv \pm m e^{{w_{1}}^{\prime}} \cosh w_{2}^{\prime} a   \hspace{1cm}  a \Delta_{2}^{\prime} = {p^{4}}^{\prime} t_{2} \hspace{1cm} x_{2}^{4} = t_{2}, 
\end{equation}
$a$ is real, $m$ is a positive constant, and the $p^{k}$ are as-yet-unknown functions of $w_{i}$ and $n^{k}.$ In (\ref{QEDnu3}), the integral over $x^{\mu}$ is a surface integral over the surface $S$ which in the frame for (\ref{K(2,1)psiab}) is just $x^{4}$ = 0. In the same frame $N^{\mu}$ = $\{0,0,0,1\}.$

	{\textit{Matrices}}. At this point in the process, we look to combine projection operators so that the matrices have a space-time invariant form. One can show that none of the four projection operators is space-time symmetric on its own. Hence, we require that two projection operators combine to make a space-time invariant matrix. In particular, we assume that
$$m(i i^{\dagger} + j j^{\dagger}) \gamma^{4} = + (E \gamma^{4} - p^{k} \gamma^{k} + M \cdot 1) \hspace{6cm}  $$
\begin{equation} \label{Epmatrix3}
 m(k k^{\dagger} + l l^{\dagger}) \gamma^{4} = - (E^{\prime} \gamma^{4} - {p^{k}}^{\prime} \gamma^{k} + M^{\prime} \cdot 1),    \hspace{1cm} i,j,k,l \in \{ u_{a}^{+},  u_{b}^{-},  v_{a}^{+},  v_{b}^{-} \},
\end{equation}
where we can use the $\theta$ = 0 basis because the phase factors would cancel anyway. The parameters $m,$ $M,$ and $M^{\prime}$ are positive or zero. The choice of signs on the right in (\ref{Epmatrix3}) is made for convenience. We assume that $\{p^{k},E\}$ and $\{{p^{k}}^{\prime},E^{\prime}\}$ transform as 4-vectors in order to have energy-momentum $p_{\mu}$ connect the invariant matrix $p_{\mu} \gamma^{\mu}$ with the invariant phase $p_{\mu} x^{\mu}$ in (\ref{QEDnu3}). 

	One purpose of developing a characterization of spacetime objects is to provide a format for investigating alternatives. Such considerations have their place elsewhere. Hence ways other than (\ref{Epmatrix3}) to make an invariant matrix are to be considered elsewhere. 

	There are just three ways to select two sets of two pairs $\{i,j\}$ and $\{k,l\}$ from a set of four. Hence we consider three cases.

	Case 1: $i$ = $u^{+}_{a},$ $j$ = $u^{-}_{b},$ $k$ = $ v_{a}^{+},$ and $l$ = $ v_{b}^{-}.$ By (\ref{abu+u-1}), (\ref{uvabcd2}), and the $ij$ part of (\ref{Epmatrix3}), we get 
\begin{equation}	\label{case1}
 a =  u^{+}  \hspace{1cm} b = u^{-}, 
\end{equation}
$$m e^{w_{1}} = m e^{w_{3}} =  M \hspace{1cm} w_{2} = w_{4} $$ $$ E = M \cosh w_{2} \hspace{1cm} p^{k} =  M \sinh (w_{2}) n^{k} \hspace{1cm}  {\mathrm{(Case}} \hspace{0.3cm} {\mathrm{1)}}. $$

	By (\ref{case1}), the expressions (\ref{uvabcd2}) for $v_{a}^{+}$ and $ v_{b}^{-}$ simplify. Then the $kl$ part of (\ref{Epmatrix3}) gives
\begin{equation}	\label{case1b}
 m e^{w_{5}} = m e^{w_{6}} =  M^{\prime} \hspace{0.5cm}  E^{\prime} = -  M^{\prime} \cosh w_{2} \hspace{0.5cm} {p^{k}}^{\prime} =  M^{\prime} \sinh (w_{2}) n^{k} .
\end{equation}

	Comparing (\ref{case1}) and (\ref{case1b}), we see that the masses $M $ and $M^{\prime}$ may differ. This is a normalization problem that can be fixed by making the sum of the projection matrices proportional to unity. By (\ref{Epmatrix3}), one can show that $M = M^{\prime}$ if and only if
\begin{equation}	\label{case1c}
 u^{+}_{a} {u^{+}_{a}}^{\dagger} + u^{-}_{b} {u^{-}_{b}}^{\dagger} + v^{+}_{a} {v^{+}_{a}}^{\dagger} + v^{-}_{b} {v^{-}_{b}}^{\dagger} \propto 1,
\end{equation}
where $1$ stands for the unit 4$\times$4 matrix.

	The normalization values for $u^{+}_{a}$ and $ u^{-}_{b}$ must be the same to make a spacetime invariant, i.e. by (\ref{uvortho}) and (\ref{case1}) we have $e^{w_{1}} \cosh w_{2}$ = $e^{w_{3}} \cosh w_{4},$ and, by (\ref{case1b}), the normalization values for $v^{+}_{a}$ and $ v^{-}_{b}$ must be equal. But the normalizations for all four are the same only for equal particle and antiparticle masses $M$ = $M^{\prime}.$

	By (\ref{case1}), (\ref{case1b}), and (\ref{case1c}), the eigenvector basis (\ref{uvabcd2}) reduces to 
$$u^{+}_{a} = \sqrt{\frac{ M}{m}} \pmatrix{  e^{w_{2}/2} u^{+} \cr e^{-w_{2}/2} u^{+} } \hspace{0.5cm} u^{-}_{b} = \sqrt{\frac{ M}{m}}  \pmatrix{  e^{-w_{2}/2} u^{-} \cr e^{w_{2}/2} u^{-} }  $$
\begin{equation}	\label{case1d} 
v^{+}_{a}= e^{i\alpha} \sqrt{\frac{ M}{m}}  \pmatrix{  e^{-w_{2}/2} u^{+} \cr  - e^{w_{2}/2} u^{+} } \hspace{0.5cm} v^{-}_{b} = e^{i \delta} \sqrt{\frac{ M}{m}}  \pmatrix{ -  e^{ w_{2}/2} u^{-} \cr  e^{-w_{2}/2} u^{-} },
\end{equation}
Thus the bases found for Case 1 are all essentially the same as the $uv$ basis of Part I, I(9).  We recover the $uv$ basis when we select 
\begin{equation}	\label{case1e}
  m = M  \hspace{1cm} w_{2} = w \hspace{1cm}\alpha = \delta = 0  \hspace{1cm}  {\mathrm{(Case}} \hspace{0.3cm} {\mathrm{1)}}. 
\end{equation}
By Part I, we conclude that Case 1 gives propagators that can describe a free electron and a free positron.

	Case 2: $i$ = $ u_{a}^{+},$ $j$ = $ v_{a}^{+},$ $k$ = $ u_{b}^{-},$ and $l$ = $ v_{b}^{-}.$ By (\ref{abu+u-1}), (\ref{uvabcd2}), and the $ij$ expression in (\ref{Epmatrix3}), we get 
\begin{equation}	\label{case2} 
  a   = e^{i\nu} u^{-}  \hspace{1cm} b = e^{i \zeta} u^{+} \hspace{1cm}  {\mathrm{(Case}} \hspace{0.3cm} {\mathrm{2)}}
\end{equation}
$$M = 0 \hspace{1cm} w_{1} = w_{5} \hspace{1cm} E = m e^{w_{1}} \cosh w_{2} \hspace{1cm}  p^{k} = m e^{w_{1}}  \cosh(w_{2}) n^{k} , $$
where $\nu$ and $\zeta$ are arbitrary phases. By (\ref{case2}), the $kl$ expression in (\ref{Epmatrix3}) gives
\begin{equation}	\label{case2a} 
 M^{\prime} = 0 \hspace{0.5cm} w_{3} = w_{6} \hspace{0.5cm} E^{\prime} = - m e^{w_{3}} \cosh w_{4} \hspace{0.5cm}  p^{k} = m e^{w_{3}}  \cosh(w_{4}) n^{k} .
\end{equation}

	By (\ref{case2}) and (\ref{case2a}) we see that spacetime invariance allows particle and antiparticle to have different energy-momenta when they are in one of the states of this basis. When we require the sum of the projection operators to be a multiple of the unit matrix, (\ref{case1c}), we have
\begin{equation}	\label{case2b}
E = - E^{\prime} \hspace{0.5cm}  p^{k} = {p^{k}}^{\prime} \hspace{0.5cm} e^{w_{3}}  \cosh(w_{4}) = e^{w_{1}}  \cosh(w_{2}) .
\end{equation}
For simplicity, we take $w_{3}$ = $w_{1}$ and $w_{4}$ = $w_{2}.$

	By (\ref{case2}), (\ref{case2a}), and (\ref{case2b}), the eigenvector basis (\ref{uvabcd2}) reduces to
$$u^{+}_{a} = e^{w_{1}/2}\pmatrix{ e^{w_{2}/2} u^{+} \cr e^{i \nu} e^{-w_{2}/2} u^{-} } \hspace{0.5cm} u^{-}_{b} = e^{w_{1}/2} \pmatrix{ e^{-w_{2}/2} u^{-} \cr e^{i \zeta} e^{w_{2}/2} u^{+} } $$
\begin{equation}	\label{case2c} 
v^{+}_{a}= e^{i\alpha} e^{w_{1}/2} \pmatrix{ e^{-w_{2}/2} u^{+} \cr  - e^{i\nu} e^{w_{2}/2} u^{-} } \hspace{0.5cm} v^{-}_{b} = e^{i \delta} e^{ w_{1}/2} \pmatrix{ - e^{ w_{2}/2} u^{-} \cr e^{i \zeta} e^{-w_{2}/2} u^{+} } ,
\end{equation}

	The bases that have spacetime symmetry under Case 2 are equivalent to the $fg$ basis II(1). To duplicate those particular pairs we choose 
\begin{equation}	\label{fg2}
w_{1} = 0 \hspace{1cm} w_{2} = w \hspace{1cm} \alpha = \delta = \nu = \zeta = 0 . \hspace{0.5cm}  {\mathrm{(Case}} \hspace{0.3cm} {\mathrm{2)}}
\end{equation}
The normalization parameter $w_{1}$ can be absorbed in $m$ and $\alpha,$ $\delta,$ $\nu,$ and $\zeta$ are arbitrary phases. Hence, by Part II, Case 2 gives bases that lead to the neutrino and antineutrino propagators II(17) and II(21).

	Case 3: $i$ = $ u_{a}^{+},$ $j$ = $ v_{b}^{-},$ $k$ = $ u_{b}^{-},$ and $l$ = $ v_{a}^{+}.$ By (\ref{abu+u-1}), (\ref{uvabcd2}), and the $ij$ part of (\ref{Epmatrix3}), we get 
\begin{equation}	\label{case3} 
 a = u^{+}  \hspace{1cm} b = - u^{-}, \hspace{1cm}  {\mathrm{(Case}} \hspace{0.3cm} {\mathrm{3)}}
\end{equation}
$$m e^{w_{1}} = m e^{w_{6}} = M \hspace{1cm} w_{2} = - w_{4}   $$ $$ E = M \cosh w_{2} \hspace{1cm} p^{k} = M \sinh (w_{2}) n^{k} . $$
By (\ref{case3}), the expressions (\ref{uvabcd2}) for $u_{b}^{-}$ and $ v_{a}^{+}$ simplify. Then the $kl$ version of (\ref{Epmatrix3}) yields
\begin{equation}	\label{case3a} 
 m e^{w_{3}} = M^{\prime}  \hspace{0.5cm} w_{3} = w_{5} \hspace{0.5cm} E^{\prime} = - M^{\prime} \cosh w_{2} \hspace{0.5cm}  {p^{k}}^{\prime} = M^{\prime}  \sinh(w_{2}) n^{k},
\end{equation}
Requiring that the sum of the projection matrices be proportional to unity, (\ref{case1c}), gives 
\begin{equation}	\label{case3b}
E = - E^{\prime}  \hspace{1cm} p^{k} = {p^{k}}^{\prime}   \hspace{1cm} M = M^{\prime} .
\end{equation}

		By (\ref{case3}), (\ref{case3a}), and (\ref{case3b}), the eigenvector basis (\ref{uvabcd2}) reduces to
$$u^{+}_{a} = \sqrt{\frac{ M}{m}} \pmatrix{ e^{w_{2}/2} u^{+} \cr e^{-w_{2}/2} u^{+} } \hspace{0.5cm} u^{-}_{b} = \sqrt{\frac{ M}{m}} \pmatrix{ e^{w_{2}/2} u^{-} \cr - e^{-w_{2}/2} u^{-} }  $$
\begin{equation}	\label{case3c} 
v^{+}_{a}= e^{i\alpha} \sqrt{\frac{ M}{m}} \pmatrix{ e^{-w_{2}/2} u^{+} \cr  - e^{w_{2}/2} u^{+} } \hspace{0.5cm} v^{-}_{b} = e^{i \delta} \sqrt{\frac{ M}{m}} \pmatrix{ - e^{ - w_{2}/2} u^{-} \cr  - e^{w_{2}/2} u^{-} },
\end{equation}
The basis (\ref{case3c}) is essentially the same as the $uv$ basis in Part I, I(9). To get the $uv$ basis within a couple of signs, $u_{b}^{-} \rightarrow$ $- v_{-}^{-}$ and $v_{b}^{-} \rightarrow$ $- u_{-}^{-},$ put
\begin{equation}	\label{case3d}
 m = M  \hspace{1cm} w_{2} = w \hspace{1cm}\alpha = \delta = 0 .  \hspace{0.5cm}  {\mathrm{(Case}} \hspace{0.3cm} {\mathrm{3)}}
\end{equation}
Thus, by Part I, Case 3 yields a basis that gives a free electron and a free positron propagator.

	Spacetime invariance, i.e. the requirement (\ref{Epmatrix3}), has returned only eigenvector bases that give electron or neutrino-like propagators. It follows that the only space-time invariant spin 1/2 propagators are electron-like or neutrino-like, within the limits of the assumptions applied in the process developed in this series of papers.

	Furthermore spacetime symmetry results only when the two planes of an eigenvector basis are complementary in $E^{4}.$ 
Hence we have found a geometric property that signals the ability to construct spacetime. 

\pagebreak
	
\appendix

\section{U(1)$\times$SU(2) Theorem} 

	Let $u^{+}$ and $u^{-}$ be orthonormal 2-vectors. Let $a$ and $b$ be a second set of orthonormal 2-vectors, as in (\ref{normal0}) and (\ref{normal2}). As shown in Problem 2, it follows that $u^{+}$ and $u^{-}$ are eigenvectors for rotations in one  plane, plane 1, and $a$ and $b$ are eigenvectors in some plane, plane 2.

	Theorem: One orthonormal 2-vector set $\{a,b\}$ can be found from another orthonormal set $\{u^{+},u^{-}\}$ by multiplication with a rotation matrix like $R$, I(1), and a phase factor, i.e. by the action of a U(1)$\times$SU(2) transformation. The proof follows.

	Since it takes two linearly independent 2-vectors to span the space of 2-vectors, $a$ and $b$ can be expressed in terms of $u^{+}$ and $u^{-}.$ Let $T$ be the matrix of coefficients relating $a$ and $b$ to $u^{+}$ and $u^{-}.$
\begin{equation}	\label{T1}
 \pmatrix{ a  \cr b} = \pmatrix{T_{11} && T_{12} \cr T_{21} && T_{22}}  \pmatrix{ u^{+}  \cr u^{-}} .
\end{equation}
We need to show that $T$ has the form of a phase factor times a matrix like $R$, I(1). 

	Since $a^{\dagger} a$ = 1, $b^{\dagger} b$ = 1, and $a^{\dagger} b$ = 0, we get
\begin{equation}	\label{T1a}
 \mid T_{11}\mid^{2} + \mid T_{12}\mid^{2} = 1 \hspace{1cm} \mid T_{21}\mid^{2} + \mid T_{22}\mid^{2} = 1 \hspace{1cm} T_{11}^{\ast} T_{21} = - T_{12}^{\ast} T_{22} 
\end{equation}
Write $(a^{\dagger} b)^{\ast}(a^{\dagger} b)$ = 0 in terms of the $T_{ij}$s. We have
\begin{equation}	\label{T1b}
 \mid T_{11}\mid^{2} \mid T_{21}\mid^{2} + T_{11}^{\ast} T_{21} T_{12} T_{22}^{\ast} + T_{11} T_{21}^{\ast} T_{12}^{\ast} T_{22} + \mid T_{12}\mid^{2} \mid T_{22}\mid^{2}  = 0 . 
\end{equation}
Now use (\ref{T1a}) to eliminate $T_{11}$ and $T_{11}^{\ast}$ from the first three terms in (\ref{T1b}). Note that it is not necessary to divide by any $T_{ij}$ components. By simplifying the result we get 
\begin{equation}	\label{T1c}
 \mid T_{21}\mid^{2} = \mid T_{12}\mid^{2} . 
\end{equation}
By (\ref{T1a}) and (\ref{T1c}), we get
\begin{equation}	\label{T1d}
 \mid T_{11}\mid^{2} = \mid T_{22}\mid^{2} . 
\end{equation}
These equations force $T$ to take the following form in terms of real valued phases $\alpha,$ $\beta,$ $\gamma,$ $\delta,$ and $\eta.$ We get
\begin{equation}	\label{T2}
 \pmatrix{T_{11} && T_{12} \cr T_{21} && T_{22}}  = \pmatrix{ \exp{i\alpha} \cos{\eta} && \exp{i\beta} \sin{\eta} \cr -\exp{i\gamma} \sin{\eta} && \exp{i\delta} \cos{\eta}},
\end{equation}
where the new parameters here on the right have no relation to the other uses of the same symbols elsewhere in this paper.

	By the rightmost equation in (\ref{T1a}), i.e. $a^{\dagger} b$ = 0, we find that $\beta + \gamma$ = $\alpha + \delta + 2 n \pi.$ Since $\alpha$ = $(\alpha + \delta)/2$ $+ (\alpha-\delta)/2,$ $\delta$ = $(\alpha + \delta)/2$ $- (\alpha-\delta)/2,$ $\beta$ = $(\alpha + \delta)/2$ $+ (\beta-\gamma)/2 + n \pi,$ $\gamma$ = $(\alpha + \delta)/2$ $- (\beta-\gamma)/2 = n \pi,$ we have 
\begin{equation}	\label{T3}
 \pmatrix{T_{11} && T_{12} \cr T_{21} && T_{22}}  = e^{i(\alpha + \delta)/2}\pmatrix{ \exp{[i(\alpha-\delta)/2]} \cos{\eta} && (-1)^{n} \exp{[i(\beta-\gamma)/2]} \sin{\eta} \cr (-1)^{n+1}\exp{ [-i(\beta-\gamma)/2]} \sin{\eta} && \exp{[-i(\alpha-\delta)/2]} \cos{\eta}}.
\end{equation}
We see that $T$ is a phase factor times a unimodular unitary matrix; the set of such matrices forms a $2 \times 2$ representation of SU(2), which is equivalent to a representation of rotations in $E^{3}$ or $E^{4}.$
 
	To put the SU(2) matrix in (\ref{T3}) in the same form as $R$, see I(1), we introduce new variables $\kappa$ and $N^{k}$, 
\begin{equation}	\label{T4}
\cos{(\kappa/2)} = \cos{[(\alpha - \delta)/2]} \cos{\eta} \hspace{1cm} N^{3} \sin{(\kappa/2)} = \sin{[(\alpha - \delta)/2]} \cos{\eta} \end{equation}
$$ N^{1} \sin{(\kappa/2)} = (-1)^{n} \sin{[(\beta - \gamma)/2]} \sin{\eta} \hspace{1cm} N^{2} \sin{(\kappa/2)} =  (-1)^{n} \cos{[(\beta - \gamma)/2]} \sin{\eta}. $$
One can show that $N^{k}$ is a unit 3-vector, i.e. $\sum {N^{k}}^2$ = 1, for $\kappa \neq$ 0. In terms of the new variables, we get
\begin{equation}	\label{abu+u-}
 \pmatrix{ a  \cr b} = T \pmatrix{ u^{+}  \cr u^{-}} = e^{i\chi} [\tau^{4} \cos (\kappa/2) + i N^{k} \tau^{k} \sin (\kappa/2)]\pmatrix{ u^{+}  \cr u^{-}} = e^{i\chi} e^{i N^{k} \tau^{k} \kappa/2} \pmatrix{ u^{+}  \cr u^{-}},
\end{equation}
where $\chi$ = $(\alpha + \delta)/2.$ The $2 \times 2$ matrices $\tau$ are the same as the $\sigma$ matrices, I(2), 
\begin{equation} \label{tau}
\tau^{1} = \pmatrix{0 && 1 \cr 1 && 0} \hspace{0.3in} \tau^{2} = \pmatrix{ 0 && -i \cr i && 0} \hspace{0.3in} \tau^{3} = \pmatrix{ 1 && 0 \cr 0 && -1} \hspace{0.3in} \tau^{4} = \pmatrix{ 1 && 0 \cr 0 && 1}.
\end{equation}
We write a spin matrix as a $\tau$ when it is applied to a pair of 2-vectors and we write the spin matrix as a $\sigma$ when it is applied to the two components of a 2-vector. 

	Comparing (\ref{abu+u-}) and I(1) completes the proof.




 \section{Problems} 

\noindent 1. Consider the pair $u_{+}^{+}$ = $\pmatrix{ e^{w/2} u^{+} \cr e^{-w/2} u^{+} }$ of the $uv$ basis, I(9). Show that $u_{+}^{+}$ does not represent any vector quantity $\chi$ in $E^{4}.$ Find two pairs $X_{1}$ and $X_{2}$ that do represent vector quantities $\chi_{1}$ and $\chi_{2}$ and whose sum is $u_{+}^{+},$ i.e. $X_{1} + X_{2}$ = $u_{+}^{+}.$ Assume that the upper eigenvector $u^{+}$ (a 2-vector) is an eigenvector for rotations about the origin in plane 1 and the lower eigenvector $u^{+}$ is an eigenvector for rotations about the origin in plane 2. Also assume that plane 1 and plane 2 span $E^{4}.$ 

[Hint: An eigenvector represents a vector perpendicular to the plane of rotation. A vector in the plane of rotation is represented by a sum of the eigenvectors for rotations in the plane.]

\vspace{0.3cm}

\noindent 2. Prove the following theorem. Theorem: two orthonormal 2-vectors, say $\delta$ and $\epsilon,$ are the eigenvectors of some rotation. (One way is identify the parameters in $u^{+}$ in I(3) in terms of the components of one 2-vector and use orthonormality to show the other must have the form of $u^{-}$ in I(3). Then construct the matrix $R$ and show $R\delta$ = $\exp{(i\theta/2)} \delta$ and that $R\epsilon$ = $\exp{(-i\theta/2)} \epsilon ,$ thereby completing the proof.)

\pagebreak

\vspace{0.3cm}

\noindent 3. Consider two coordinate planes 12 and 34, where 1234 indicates rectangular coordinates. If we decide the positive rotation direction for plane 12 then there are two ways to orient the rotations in plane 34. If one orientation gives a positive phase $+\theta/2$ for a rotation in the plane 34 then the other orientation gives a negative phase $-\theta/2.$ Thus the occurance of two orientations for rotations in plane 34 means the lower 2-vector can have either the same phase as or the opposite phase of the upper 2-vector. 

	This suggests that we investigate both combinations of phases, $++$ and $+-,$ for upper and lower 2-vectors. The rotated basis pairs for an electron in Part I and a neutrino in Part II have upper and lower 2-vectors with like phase $\theta/2.$ Carry through the processes from rotated basis to spacetime invariant electron and neutrino propagators with the upper 2-vectors having phase $\theta/2$ and the lower 2-vectors having the opposite phase $-\theta/2.$

\vspace{0.3cm}

\noindent 4. Show that $2\times2$ matrices cannot provide a nontrivial representation of all rotations about the origin in Euclidean spaces of five or more dimensions. One way to show this is to consider a $2\times2$ matrix representation in a four dimensional subspace and then try to extend that representation to rotations in unrepresented planes. (For an application of $2\times2$ matrix representations of rotations in 4d subspaces of 16d, see \cite{16d}.)


\begin{thebibliography}{9}

\bibitem{partI} Shurtleff, R., hep-th/0007232.

\bibitem{partII} Shurtleff, R., hep-th/0007245.

\bibitem{16d} Shurtleff, R., quant-ph/9907012.

\end{thebibliography}
\end{document}